Thomas A. Carlson, Tijl Grootswagers, Amanda K. Robinson;

Title: An introduction to time-resolved decoding analysis for M/EEG

Thomas A. Carlson, School of Psychology, University of Sydney, NSW Australia
Tijl Grootswagers, School of Psychology, University of Sydney, NSW Australia
Amanda K. Robinson, School of Psychology, University of Sydney, NSW Australia

Abbreviated title: Time-resolved decoding analysis for M/EEG time series

Corresponding author:
Thomas A. Carlson
School of Psychology
318 Griffith-Taylor Building
University of Sydney
NSW Australia

Phone: +61 4 1449 1771
Email: Thomas.Carlson@Sydney.edu.au

Word count: 4737
Number of figures: 2

Acknowledgments: We would like to thank the reviewers and Nick McNair, Denise Moerel, Selene Petit for their feedback and suggestions.

Index keywords: Electroencephalography, magnetoencephalography, multivariate pattern analysis, brain decoding, representational dynamics

**Abstract**

The human brain is constantly processing and integrating information in order to make decisions and interact with the world, for tasks from recognizing a familiar face to playing a game of tennis. These complex cognitive processes require communication between large populations of neurons. The non-invasive neuroimaging methods of electroencephalography (EEG) and magnetoencephalography (MEG) provide population measures of neural activity with millisecond precision that allow us to study the temporal dynamics of cognitive processes. However, multi-sensor M/EEG data is inherently high dimensional, making it difficult to parse important signal from noise. Multivariate pattern analysis (MVPA) or "decoding" methods offer vast potential for understanding high-dimensional M/EEG neural data. MVPA can be used to distinguish between different conditions and map the time courses of various neural processes, from basic sensory processing to high-level cognitive processes. In this chapter, we discuss the practical aspects of performing decoding analyses on M/EEG data as well as the limitations of the method, and then we discuss some applications for understanding representational dynamics in the human brain.

**Introduction**

One of the most remarkable aspects of the human brain is its speed and processing efficiency. A familiar friend is recognized in an instant, speech communication feels like a natural dynamic exchange, and we can make split-second decisions in life-threatening situations. A crucial question in cognitive neuroscience is how the brain manages such complex tasks with ease. As non-invasive brain imaging techniques with millisecond resolution, magnetoencephalography and electroencephalography (M/EEG) offer unparalleled potential in capturing the dynamics of human cognition. Multivariate pattern analysis (MVPA or "decoding") methods, in conjunction with M/EEG data, give insight into the temporal dynamics of information processing in the brain. Not only do such methods yield insight into the time course of specific cognitive processes, but also into the relative order of different cognitive processes. This can lead to insights into the temporal dynamics of brain representations, for example how low-level visual representations are transformed into high-level object representations (Contini et al., 2017).

When MVPA was first introduced to functional magnetic resonance imaging (fMRI), it gave cognitive neuroscientists the unprecedented capacity to decode information in the brain (for recent reviews see Haxby, 2012; Haynes, 2015; Tong and Pratte, 2012). MVPA in combination with fMRI's excellent spatial resolution enabled researchers to target and scrutinize the information represented in different brain areas. MVPA methods for M/EEG data complement fMRI findings by providing valuable insight into the time course of neural processing. Accordingly, M/EEG decoding methods are becoming increasingly popular in the cognitive neuroscience community.

Most cognitive neuroscientists are aware of decoding methods as they have been actively used in fMRI research for two decades. Interestingly, methods for decoding mental states from the brain predates this work by a quarter century. Vidal first developed these methods for EEG in the context of brain-computer communication asking the provocative question, "can observable electrical brain signals be put to work as carriers of information in man-computer communication or for the purpose of controlling such external apparatus as prosthetic devices or spaceships?" (Vidal, 1973, pg 157). Vidal's and subsequent work focused on the practical applications of having access to the brain's internal mental states. There are many useful applications for decoding methods in M/EEG research, such as Brain-Computer Interfaces (BCI; e.g., Wolpaw et al., 2002), lie detection (e.g., Davatzikos et al., 2005) and the diagnosis of brain disorders (e.g., Ewers et al., 2011). Practical applications, such as these, place greater weight on performance over explanation. If diagnostic accuracy for Schizophrenia using a particular method is increased by 2%, this is a significant achievement that can have wide-reaching benefits. In contrast, the cognitive neuroscientist is interested in how the brain works. Here, the priority is to elucidate neural mechanisms underlying cognitive processes, and metrics like prediction accuracy are of lesser importance (Hebart and Baker, 2017). It is the cognitive neuroscience application of decoding methods for M/EEG that is the topic of this chapter. Henceforth, we use the term decoding to reference the use of this method for understanding the brain.

This chapter focuses explicitly on time-resolved decoding analysis for M/EEG. There are a variety of domains in which MVPA can be applied to M/EEG data (frequency analysis, wavelet decomposition, connectivity, etc.). We will limit the scope of our discussion to time-series decoding, but many of the same principles apply to other domains. This research uses paradigms that closely resemble EEG's event-related potential (ERP) or MEG's event-related field (ERF) research. In these experiments, participants are presented with stimuli (e.g., images, sounds, etc.) from different conditions (e.g., faces vs houses; attended vs unattended) and the analysis is time-locked to the presentation of the stimulus. MVPA and ERP/ERF analyses both seek to determine if there are differences between experimental conditions. The main difference is that ERP/ERF research uses a univariate approach, such as testing for differences between conditions from a single recording site or from data averaged across multiple sites. MVPA, in contrast, utilizes pattern information across recording sites and thus is often a more sensitive measure. As the topic of this chapter is MVPA, we will not discuss further differences between the approaches except for illustrative points (for more information see Grootswagers et al., 2017b; Hebart and Baker, 2017). For those interested in ERP/ERF analysis or more general information about the analysis of EEG recordings, we refer the reader to the excellent text on ERP research by Luck ((Luck, 2005).

The fundamental goal of the decoding analysis is to learn what and how information is represented by the brain (see also Chapters by King et al. and by Diedrichsen, this volume). M/EEG recordings are a measure of brain activity with high temporal resolution, and MVPA is a sensitive measure to determine whether or not there is information in the recordings that can distinguish between experimental conditions. The combination of the two can tell the researcher if there is information in the recordings of brain activity that can distinguish experimental conditions for each point in time. Going beyond this to address the more critical questions of "what" and "how" information is represented in the brain requires specific knowledge of the field of inquiry, which requires the development of robust experimental paradigms.

The goal of this chapter is to inform readers with little MVPA experience about the mechanics of applying these methods to M/EEG data and give guidance on getting at the critical questions of

"what" and "how" information is represented in the brain using these methods. We assume the reader has basic knowledge M/EEG and thus will focus on the practical details of decoding analyses rather than aspects of conducting M/EEG experiments. We also note that M/EEG-MVPA is a relatively new field of research. On the one hand, this means that many exciting new theoretical questions can be explored using these powerful analysis methods. On the other hand, it means the techniques themselves are still evolving, and there is currently limited technical guidance (see Grootswagers et al., 2017b for an advanced tutorial ). At times, we will give recommendations based on our own experience that lack a scientific reference. We expect many of these recommendations will be followed up empirically in the coming years.

In this chapter, we first provide a foundation by situating pattern analysis methods in the context of M/EEG data. We then discuss some of the intricacies of using decoding methods for analyzing M/EEG data. We then conclude with a section describing advanced methods to familiarize the reader with more sophisticated approaches to answering the questions of "what" and "how" information is represented in the brain.

**The basics of time-resolved decoding analysis for M/EEG**

To give intuition to the analysis and a framework for discussion, we describe a hypothetical experiment. Consider a standard EEG system with 64 sensors (Figure 1A), each of which continuously measures local electrical currents on the scalp (Figure 1B). In each trial, participants are shown an image of either an 'X' or an 'O', and there are 40 trials for each condition. To simplify the description of the analysis, we start by considering only two electrodes centered on occipital cortex (e.g., O1 and O2) and data from a single time point (e.g., 100 ms post stimulus onset). This narrow slice of data can be represented as a 2D scatterplot (Figure 1C). In the plot, the two axes are the measurements from the two electrodes in microvolts (μV) for the single time point. The 80 data points are the measurements recorded for the individual trials, denoted by 'X's and 'O's. In this simplified scenario, the EEG data takes the form of the familiar diagram given in virtually all texts describing pattern classification analysis (e.g., Duda et al., 2001).

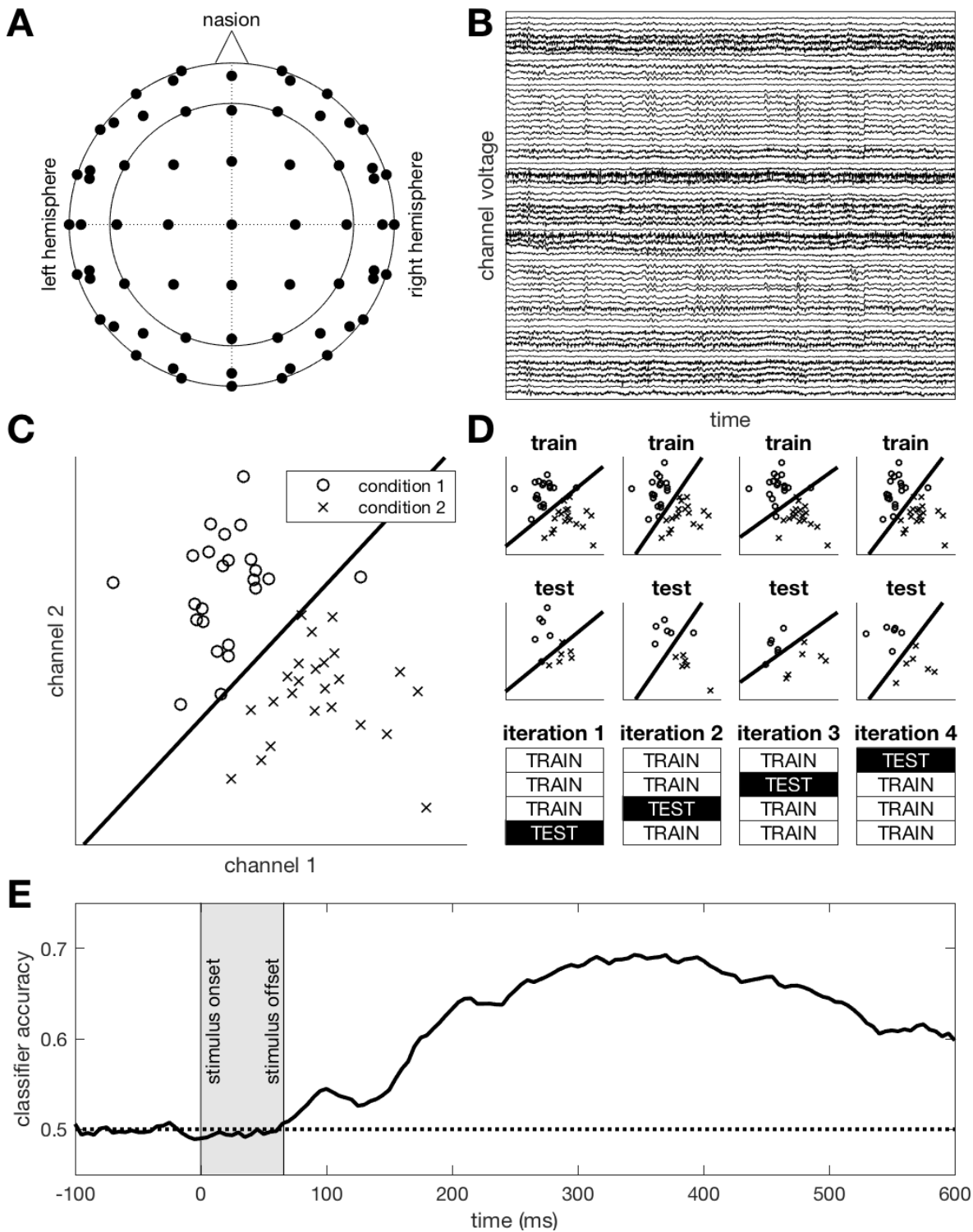

**Figure 1. Time-resolved decoding analysis for M/EEG. (A)** Top down view of a channel layout for a 64-channel EEG. **(B)** Sample recording data from an EEG experiment. **(C)** Scatterplot showing hypothetical EEG data from a single time point for a decoding experiment. The two axes are the measured current from two EEG electrodes. The points are individual trials for the 'X' and 'O' conditions. The line denotes the optimal decision boundary for classifying 'X's and 'O's derived from the generative model. **(D)** Four iterations of cross-validation using 75% of the data to train the classifier and 25% to test the classifier. The top row of plots show the training data along with the classification boundary computed using Linear Discriminant Analysis (LDA) for each iteration. The plots below show the test data for each iteration with the decision boundary computed from the training data. **(E)** An example of a time-resolved decoding analysis showing classifier accuracy averaged over participants (data from Grootswagers et al., 2017a).

Pattern classification analysis is used to determine if there is information in the EEG measurements that can predict whether the participant was viewing an 'X' or an 'O' on a given trial. The analysis is a two-step process. The first step is to train the classifier to find a decision boundary that best separates the 'X's and 'O's using the EEG data. For purposes of this initial description, we use a linear classifier, which uses a linear decision boundary for classification. In the plot in Figure 1C, the observations to the left of the boundary are classified as 'O's and those to the right of the boundary are classified as 'X's.

The second step is testing the classifier, where a trained classifier is used to predict whether the participant was viewing an 'X' or an 'O' on a new set of trials. It is essential to use independent data for the two steps (training and test), to show the classifier can generalize to independent data outside the training sample. For this, we use cross-validation, which involves splitting the data into training and test sets. In this example, we divide the data into 4 blocks and use 75% of the data to train the classifier and the 25% to test the classifier. The classifier is trained four times, each time using three blocks of data to train the classifier and the remaining block to test the classifier (Figure 1D). The average prediction accuracy (performance) of the classifier across the four iterations is taken as an estimate of decoding performance for classifying 'X's and 'O's from EEG data at this timepoint. Importantly, if the classifier accuracy is above chance (i.e., 50%), this is taken as evidence there were different patterns of activation across the two channels when the participant viewed an 'X' or an 'O'.

To bring this simplified scenario back to a full M/EEG dataset, it is only a matter of scaling up. First, instead of limiting the analysis to two channels, all of the channels or a selected subset of the channels is used for the analysis. Second, the analysis is repeated for every time point instead of running the analysis on only a single time point. The result is a time-varying measure of stimulus/condition decodability (Figure 1E). Finally, the experiment is repeated in multiple subjects to enable statistical inference treating subject as a random effect (see section on Statistical assessment).

The above description provides a broad overview of the critical features of decoding analysis for M/EEG. In the following sections, we go into more detail about the steps in the analysis and discuss specific issues for the analysis of M/EEG data.

**Practical aspects of M/EEG decoding analysis**

***On preprocessing M/EEG data for MVPA***

M/EEG data is inherently noisy, picking up electrical activity from a variety of non-brain sources such as muscle activity, eye movements, and environmental noise. M/EEG research uses many preprocessing procedures to reduce this noise, such as filtering, resampling, and artefact rejection. The goal of preprocessing is to increase the signal-to-noise ratio (SNR), reducing the probability of obtaining a false negative (i.e., failing to observe an effect that is present). Preprocessing pipelines can vary between different laboratories (and even across individual researchers in a laboratory), as there is no consensus on preprocessing pipelines. The variability in preprocessing pipelines has been identified as an important issue for reproducibility in neuroimaging research (Poldrack et al., 2017). In developing a preprocessing pipeline, one thus needs to balance the desire to maximize SNR with minimizing the number of preprocessing steps to ensure the results are robust.

The analysis of ERP/ERF in EEG/MEG respectively has been around for decades, while decoding methods are relatively new. A natural starting point for thinking about a preprocessing pipeline for M/EEG decoding analysis might be to adopt established methods from ERP/ERF research (Luck, 2005). However, MVPA is different to ERP/ERF analyses, thus the preprocessing steps need not be the same. In particular, decoding is more robust to artefacts than these other methods. Standard classifiers, such as linear discriminant analysis (LDA) and support vector machines (SVM), implicitly model the noise in the data. Therefore, classifiers can replace steps that typically require laborious manual inspection such as bad sensor and artefact rejection. To understand this, we need to look at how machine learning classifiers like LDA and SVM maximize the prediction.

First, the classifier assigns weights to each sensor. High weights enhance information in the measurement that informs the classification, and low weights suppress uninformative information in the measurement. This aspect is relevant because it potentially makes some preprocessing procedures redundant. In the context of EEG, an example is a broken or very low impendence electrode. In a standard ERP analysis, this electrode would be removed or interpolated during preprocessing. For MVPA, this step can be omitted because the classifier would learn in training the noisy electrode is not informative for the prediction and assign it a low weight. Note also that channel interpolation is a linear combination of surrounding electrodes and thus adds no additional information to the classifier. Eye-blink and muscular artefacts are other examples. For ERP analysis, these artefacts can have a significant effect on the quality of the data and thus need to be removed. For MVPA, these artefacts are of lesser concern. If the artefact is not informative for the prediction, the classifier will suppress the artefacts by giving low weight to the component of the signal related to the artefacts. Removing trials with eye-blink artefacts is thus another preprocessing step that potentially could be removed from the MVPA preprocessing pipeline (Grootswagers et al., 2017b). Note that artefacts confounded with the conditions of interest are a serious concern for all (MVPA and ERPs) analyses!

Third, classifiers can cancel noise in the data to improve the prediction. This is relevant when considering the effect of environmental noise. For ERP analysis, a noisy channel increases variance in the estimate of the average evoked response, and thus it is advisable to interpolate noisy channels or exclude them from the analysis. In contrast, classifiers (e.g., LDA) can use the information about the noise to improve prediction. For example, suppose we have two hypothetical channels. One channel contains brain activity differentiating the two experimental conditions and also environmental noise. The second channel is far from the source of brain activity and only contains the environmental noise. The classifier can use an estimate of the environmental noise from the second channel and subtract it from the first channel so that what remains is the signal that differentiates the two experimental conditions. Including the noisy channel can therefore give the classifier *more* information about the signal of interest than if the channel had been removed (Averbeck et al., 2006; Haufe et al., 2014).

In summary, MVPA is a relatively new approach to analyzing M/EEG data that does not have stringent noise reduction requirements as in typical ERP analyses. The critical lesson to be taken concerning preprocessing is that MVPA can inherently deal with some noise and artefacts in the data. This is an advantage of the decoding approach because it reduces data processing steps relying on subjective criteria (e.g., interpolation of noisy channels identified by visual inspection). While systematic research is still required, M/EEG MVPA research will need to develop its own guidelines for preprocessing. These new standards will need to take into account how classifiers operate to strike a balance between optimizing SNR and minimizing the number of preprocessing steps and options for reproducible research (Poldrack et al., 2017; for early explorations into the effect of preprocessing pipelines on M/EEG decoding, see Grootswagers et al., 2017).

*Classifier selection*

Decoding analysis generally favors linear classifiers. As the name implies, linear classifiers use a linear boundary (e.g., see Figure 1), or a hyperplane for greater than two dimensions for classification. Compelling arguments have been made for the use of linear classifiers in fMRI research (Kriegeskorte, 2011; Misaki et al., 2010; Muller et al., 2003; Mur et al., 2009; Pereira et al., 2009; Schwarzkopf and Rees, 2011). In contrast, there has been little systematic discussion of this for M/EEG. Below we give an overview of the different classifiers available for M/EEG decoding research, and step through the fMRI community's argument for linear classifiers and discuss them in the context of M/EEG.

Although non-linear classifiers have the added capacity to fit more complex class boundaries using non-linear terms, in practice they generally perform equal or worse than their linear counterparts in neuroimaging decoding studies (Misaki et al., 2010). This is because the increased flexibility of non-linear classifiers comes at the cost of overfitting, which limits generalization performance. Furthermore, the results from non-linear classifiers are harder to interpret. As previously noted, it is also essential to keep in mind that cognitive neuroscience applications of MVPA focus on understanding the brain, where improvements in performance are of marginal value (see Introduction). A small increase in performance is a hefty price for the loss of interpretability, which we discuss below. Thus, unless there is substantial justification, linear classifiers are the preferred method for cognitive neuroscience applications of MVPA for the analysis of M/EEG data.

Linear classifiers have two interpretive advantages from a cognitive neuroscience perspective. First, linear classifiers are a biologically plausible form of "read out" (DiCarlo and Cox, 2007) Specifically, the information used by a linear classifier could also be "read out" by a single downstream neuron (Kriegeskorte, 2011; Misaki et al., 2010). In the context of M/EEG, this is less compelling. Embedded in this reasoning is that the downstream neuron has access to the same information as the classifier (Carlson et al., 2017). A single EEG channel records the aggregated activity of thousands of neurons from different brain regions, and the entire EEG cap has access to most of the neurons on the cortical surface. It is not biologically plausible that a single neuron has access to all information represented on the cortical surface.

The second interpretive advantage is that linear classifiers produce weight maps that can be visualized to gain insight into the source of decodable information (Kamitani and Tong, 2005). An fMRI-MVPA experiment studying faces, for example, might show voxels in the fusiform face area (FFA; Kanwisher et al., 1997) are given high weights. The weight maps thus can be used to infer that FFA is a strong candidate source of decodable information. Similarly, for M/EEG, the classifier weights can be projected back to the topographical map of the sensors to gain insight into the source of decodable information (but see Haufe et al., 2014 and discussion below).

The final point of discussion is the choice of classifier. In a recent study, we compared the performance of five classifiers on decoding minimally preprocessed MEG data: LDA, Gaussian Naive Bayes (GNB), Linear Support Vector Machines (SVM), Spearman's rank correlation, and

Pearson's Correlation (Grootswagers et al., 2017b). The study found that SVM, LDA, and GNB performed the best, suggesting that they are all excellent choices for M/EEG decoding studies.

*Localizing the source of decodable information in M/EEG*

Cognitive neuroscientists are often interested in both spatial (i.e., where) and temporal (i.e., when) signatures of neural processing. Recovering the underlying source(s) of neural activity is a longstanding challenge for M/EEG research. This challenge extends to M/EEG decoding research, although advanced decoding methods do offer at least one possible solution (see section on Representational Dynamics). In this discussion, it is essential to make a distinction between resolution and spatial precision. Resolution is the capacity to resolve two points in space. Using MVPA, MEG studies have shown that it may be possible to resolve activation patterns spanning V1 columns that are just a millimeter in width (Cichy et al., 2015; Wardle et al., 2016). These results highlight the remarkable sensitivity of MVPA for detecting subtle differences in patterns of activation. However, localizing the precise source of this neural information (i.e., spatial precision) remains an ongoing challenge.

Broadly, there are three methods to localize sources in the brain using MVPA. Below, we will discuss two of these methods; the third will be presented later in the chapter. The first approach is weight projection. MVPA returns both a performance metric (e.g., percent correct) and the weights used by the classifier to make the prediction. The more informative the sensor, the higher the weight. One straightforward means of identifying the source of decodable information is to plot the weights on the scalp map. When interpreting these maps, it is important to consider the definition of "informative" (c.f. de-Wit et al., 2016; Haufe et al., 2014). Previously, we discussed the three ways classifiers optimize their performance. In addition to weights being assigned to distinguish condition-specific information, weights are also used by the classifier to suppress noise. When using the weight projection method to interpret the underlying sources, it is thus essential to consider only weights that reflect the differences between conditions (Grootswagers et al., 2017b; Haufe et al., 2014). The second approach is to extract multivariate brain activity from a region of interest (ROI) for MVPA. The most straightforward variant of this is to select sensors located above the ROI, for example, using ten sensors over the occipital cortex to study early visual processing. This can also be applied in a sensor-searchlight approach (similar to fMRI, see Haynes et al., 2007; Kriegeskorte et al., 2006), where the analysis is repeated on local clusters of sensors, providing a scalp map of decoding accuracies (e.g., see Collins et al., 2018; Kaiser et al., 2016).

These two methods are subject to interpretive criticisms arising from the fact that brain activity will propagate outside the M/EEG sensors located above a source (information loss), and brain activity from nearby areas will propagate into the selected sensors (information leakage). Nevertheless, it can be a useful means for gross localization (e.g., left vs right hemisphere) in some circumstances and the limitations in its interpretability are transparent. Alternatively, the sensor level data can first be projected into source space using source reconstruction methods such as minimum norm estimate (MNE; Hamalainen and Ilmoniemi, 1994) and beamforming (Hamalainen and Ilmoniemi, 1994; Van Veen et al., 1997). Multivariate time-series data can be reconstructed using these methods using virtual voxels in an ROI, and decoding analyses can be performed for different ROIs. As with univariate analysis, the quality of the source reconstruction can be improved using anatomical MRI data from the participant and fMRI data to precisely define the ROI. Notably, these more sophisticated methods for localization are also subject to issues of information loss and leakage (Brookes et al., 2012; Gohel et al., 2018; Hipp et al., 2012; Nolte et al., 2004; Sato et al., 2018) but the algorithms will attempt to minimize their effect.

In summary, MVPA is a sensitive measure for differentiating conditions based on the evoked spatial distribution of activity from M/EEG. The MVPA approach, however, remains limited for localizing sources, which stems from the fundamental inverse problem for M/EEG.

*Statistical assessment of information at the group level*

Time-resolved MVPA tests the presence of information at the group level. That is, we want to know whether across our sample of subjects, classifier performance is higher than would be expected by chance. There are multiple proposed methods to assess this, and there is no consensus about the optimal approach. Group-level classifier accuracies can, for example, be tested against chance using a *t*-test, or non-parametric tests such as a sign-rank test (Wilcoxon, 1945) or permutation test (Oostenveld et al., 2011). Because these tests are repeated at each time point, they have to also be corrected for multiple comparisons, for example using Bonferroni, FDR (Nichols and Holmes, 2002), or cluster-based corrections (e.g., Oostenveld et al., 2011; Smith and Nichols, 2009; Stelzer et al., 2013). As with the choice of test statistic, there is no consensus on the optimal method for multiple comparisons. Thus, the choice of statistical analysis must be guided by the experimental questions (see Allefeld et al., 2016; Hebart and Baker, 2017; Thirion et al., 2015).

**Advanced methods for M/EEG decoding: Representational dynamics**

Exemplar-based decoding methods expand on categorical approaches by studying the structure of how information is represented. Figure 2A-B show the difference between the two approaches. The two figures show a reconstruction of the brain's representation of 24 objects from MEG data 140 ms after the onset of the stimulus (data from Carlson et al., 2013). The stimulus set included 12 animate objects (e.g., humans and monkeys) and 12 inanimate objects (e.g., chair and kiwi fruit). In the category decoding approach (Figure 2A), the stimuli are treated as equivalent class members (labeled 'A' = animate or 'IA' = inanimate) and the analysis is conducted using standard methods (see section on the Basics of time-resolved decoding). If the analysis shows the classifier can decode animate and inanimate objects from the MEG data, this is evidence that the brain representation of the stimuli is encoding object animacy.

Exemplar-based decoding methods study how the individual exemplars of each category are represented. Here, the decoding analysis is run for all possible pairwise combinations of exemplars–including within-category. The decodability (i.e., classification performance) of each pair is taken as a measure of "distance" between the two items in the neural representation (Walther et al., 2016). Figure 2B demonstrates this approach plotting the individual stimuli. A line connecting the object exemplars in the figure indicates the decodability of the connected pair. The length of the line is proportional to decodability, that is, stimuli that are close to one another (human and monkey face) are less decodable than stimuli that are far apart (e.g., monkey and television) at the given point in time during visual processing.

Representational similarity analysis (RSA; Kriegeskorte and Kievit, 2013; Kriegeskorte et al., 2008) is the standard framework for hypothesis testing for the exemplar-based decoding approach. In the RSA, the pairwise decodability of the stimuli from the brain recordings is encoded in a representational dissimilarity matrix (RDM; Figure 2C). The rows and columns of the matrix correspond to individual exemplars, and each cell is the neural decodability between the two exemplars. Hypotheses are tested by constructing model RDMs that make predictions about the structure of the brain representation. An animacy model, for example, predicts that animate and inanimate objects form separable clusters in the representation. Formally, the animacy model predicts that objects within the animate and inanimate object categories will be close to one another (distance = 0), and objects that span the category boundary will be far apart (distance = 1) (See Figure 2C). To test the model, the entries of the observed neural RDM are correlated with the model RDM. For the animacy model, if there is a high correlation between the neural RDM and the model, this allows us to conclude that animacy is represented in the brain.

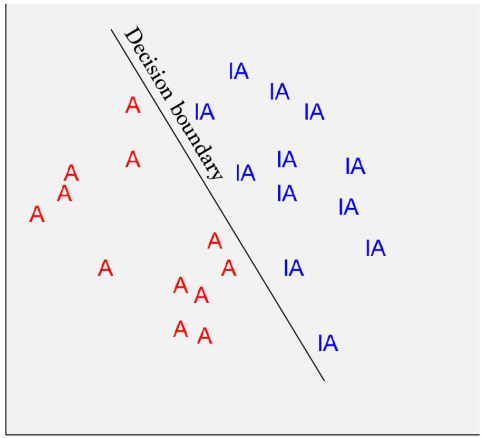
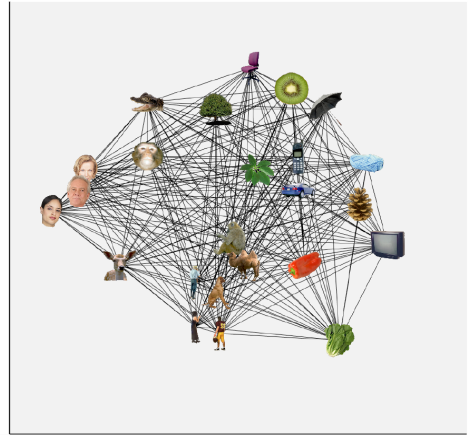
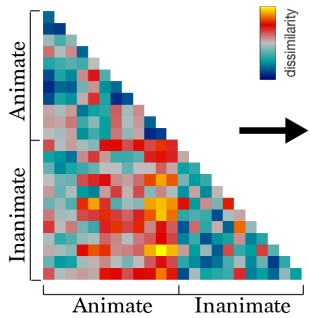
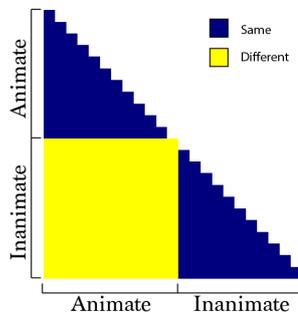
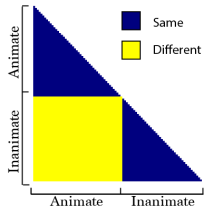
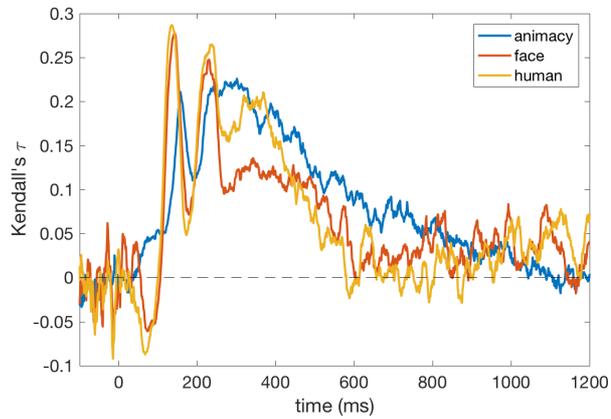
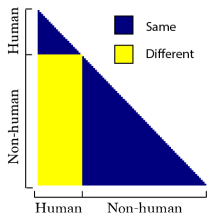
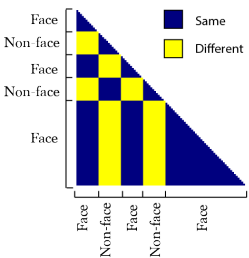
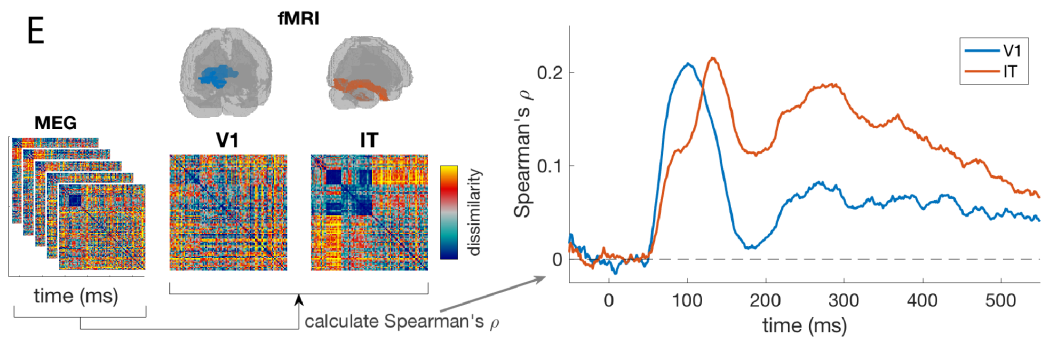

**Figure 2. Advanced methods for M/EEG decoding. (A-B)** Plots demonstrate the difference between category and exemplar-based decoding methods. Data for the plots come from a single time point (140ms post-stimulus onset) of a MEG experiment studying the brain's representation of objects (Carlson et al., 2013). **(A)** Category decoding approach with stimuli labelled by their category (A = Animate object; IA = Inanimate object) and the decision boundary from a linear classifier trained to discriminate animate and inanimate objects from the MEG data. **(B)** Exemplar decoding approach with individual stimuli displayed as images. The lines in the plot represent pairwise decodability of individual exemplars for all possible pairwise comparisons. The distance (line length) indicates the relative decodability of exemplar pairs. **(C)** Representational Similarity (RSA) model testing applied to the MEG data. The entries of the neural RDM are correlated with the animacy model to study whether the brain represents object animacy at 140 ms post-stimulus onset. **(D)** RSA used to test multiple candidate models from another MEG experiment investigating the brain's representation of objects (data from Cichy et al., 2014). Plotted is the time correlation between the time-varying RDMs from the MEG and animacy, human and face models. **(E)** Graphical depiction of the MEG-fMRI fusion approach estimating regional time-varying neural activity from a fMRI regions of interest (ROI). The time-varying MEG RDMs are correlated with RDMs from ROIs (obtained using fMRI). For each ROI, this gives a time-varying correlation indexing neural activity in the ROI. Shown is the estimated neural activity for V1 and inferior temporal cortex (data from Cichy et al., 2014).

The exemplar-based decoding approach taken in RSA makes the structure of the representation and models more explicit, which has many advantages (Kriegeskorte and Kievit, 2013; Kriegeskorte et al., 2008; Nili et al., 2014). The exemplar-based analysis can be used to study the time-varying structure of brain representations using M/EEG. As before, the analysis and model testing are performed on all of the time points. By examining how brain representations emerge over time, we gain a deeper understanding of how information is dynamically transformed and represented in the brain. Figure 2D shows the time-varying analysis from a MEG dataset with a more extensive stimulus set using several different category models (92 exemplars, Cichy et al., 2014). The figure shows that category structure for different categories (animacy, human, face) emerge at different time points in the brain's emerging representation of objects. By studying the emergence of different categories in time, studies have shown that basic level category information emerges first (e.g., human face) and abstract level category information (e.g., animacy) comes at later stages of object processing (Carlson et al., 2013; Cichy et al., 2014; Contini et al., 2017).

Exemplar-based decoding methods and the RSA framework also provide a unique solution to the challenge of acquiring data that has the spatial resolution to study regional brain activity and the temporal resolution to explore the fine grain temporal dynamics of neural activity (Cichy et al., 2014, 2016). This novel approach takes advantage of the fine-grained structure of information in brain representations, as indexed by MEG and fMRI, and integrates the data to estimate regional brain activity. The fMRI and MEG experiments are run using identical stimuli; thus, the two imaging methods produce RDMs of equal size that are directly comparable. The MEG RDM describes the brain representation at each time point, which includes activity from multiple brain regions. The fMRI RDM from an ROI describes that brain area's representation. Importantly, different brain areas have unique representations, although some areas will be similar, such as V1 and V2. By correlating a brain area's fMRI RDM with the MEG RDM for each time point, we get a time varying estimate of that area's contribution to the MEG signal across time. This method was used to study the contribution of early visual cortex (V1) and inferior temporal cortex (IT) to the MEG signal in time for object recognition (Figure 2E; Cichy et al., 2014).

**Conclusion**

M/EEG decoding methods provide a powerful set of tools for cognitive neuroscientists to gain insight into perceptual and cognitive functions by revealing the brain's processing dynamics with millisecond resolution. In this chapter, we described the fundamental aspects of M/EEG decoding analysis, practical considerations in running these analyses, and advanced methods to study representational dynamics. These methods have been incorporated into a variety of MVPA toolboxes (Bode et al., 2018; Fahrenfort et al., 2018; Gramfort et al., 2014; Hanke et al., 2009; Meyers, 2013; Oostenveld et al., 2011; Oosterhof et al., 2016). We expect M/EEG decoding to have a broad impact on the future of cognitive neuroscience research. For researchers interested in taking the next step and learning more about these methods, we refer the reader to advanced tutorials (Grootswagers et al., 2017b; Lemm et al., 2011).